\documentclass[pra,twocolumn,showpacs]{revtex4}
\usepackage{graphicx}
\usepackage{dcolumn}
\usepackage{bm}
\usepackage{amssymb}
\usepackage{amsmath}
\usepackage{epsfig}

\begin{document}

\title{Compatibility conditions from multipartite entanglement measures}
\author{Jian-Ming Cai}
\author{Zheng-Wei Zhou}
\email{zwzhou@ustc.edu.cn}
\author{Shun Zhang}
\author{Guang-Can Guo}
\affiliation{Key Laboratory of Quantum Information, University of Science and Technology
of China, Chinese Academy of Sciences, Hefei, Anhui 230026, China}

\begin{abstract}
We consider an arbitrary $d_{1}\otimes d_{2}\otimes \cdots \otimes d_{N}$
composite quantum system and find necessary conditions for general $m$-party
subsystem states to be the reduced states of a common $N $-party state.
These conditions will lead to various monogamy inequalities for bipartite
quantum entanglement and partial disorder in multipartite states. Our
results are tightly connected with the measures of multipartite entanglement.
\end{abstract}

\maketitle

\section{Introduction}

The quantum analog of the marginal distributions of a joint probability
distribution is to determine whether a given set of subsystem states $%
\left\{ \rho _{1},\rho _{2},\cdots ,\rho _{12},\cdots ,\rho
_{12\cdots j}\right\} $ comes from a single multipartite state. This
kind of compatibility problem is closely related to condensed matter
physics and chemical physics, where the significative conditions
will enable us to design powerful variational methods to
substantially simplify the computation of many physical variants.
People have utilized different ideas and techniques to attack the
tough problem of compatibility of quantum
states and obtained several important partial results \cite%
{Higuchi0,Compatible,han,Bravyi,Butterley,Higuchi0309186,Christandl0409016,Christandl0511029,Klyachko0409113,Daftuar,Jones2005,Hall2005,Hall0610031}%
. However, it is still very difficult to present compatibility conditions
for a set of general $m$-party reduced states and a multipartite state. The
difficulty comes from the phenomenon of multipartite entanglement. Since the
number of local invariants in $N$-party density matrices will increase
exponentially with the particle number $N$ \cite{Linden}, how to quantify
the measure and characterize the structure for multipartite entanglement
becomes obscurity. Nevertheless, due to the substantial significance for
distributed quantum information processing and strongly correlated physics
\cite{Entanglement and Phase Transitions,HSC}, many efforts for exploring
multipartite entanglement have been done \cite%
{Wong,Hyperdeterminants,Mayer,Gerardo,Mintert0,Barnum,ITMGE,Mintert4,Mintert1,Siewert,Lamata,Gour}

In Refs. \cite{Mintert0,Mintert1}, Mintert and co-workers\textit{\ }define
multipartite concurrence with a single factorizable observable, which can
effectively characterize quantum correlations in an $N$-party pure state.
From the information viewpoint, we propose a polynomial stochastic local
operations and classical communication (SLOCC) \cite{SLOCC} invariant $%
\mathcal{E}_{12\cdots N}$ in \cite{ITMGE}, and show that $\mathcal{E}_{1234}$
satisfies the necessary conditions for a natural entanglement measure. By
investigating the relations among different local invariants, it is possible
for us to gain insight into the nature and complex structure of multipartite
entanglement, as well as its connection with the general compatibility
problem.

In this paper, we find several necessary conditions for general $k$-party
density matrices to be the reduced states of a common arbitrary $%
d_{1}\otimes d_{2}\otimes \cdots \otimes d_{N}$ multipartite state. The main
idea is to establish a direct connection between the information-theoretic
measure of multiqubit entanglement and Mintert concurrence based on a single
factorizable physical observable. Moreover, we obtain monogamy inequalities
for bipartite quantum entanglement \cite{Osborne06} and partial disorder in
multipartite states. Our results reveal explicitly the relationship between
multipartite entanglement measures, the general compatibility problem and
the monogamous nature of entanglement.

The structure of this paper is as follows. In Sec. II we first establish the
relationship between information-theoretic entanglement measure and
multipartite concurrence, from which we derive compatibility conditions for
general $d_{1}\otimes d_{2}\otimes \cdots \otimes d_{N}$ multipartite states
in Sec. III . As applications of these compatibility conditions, in Sec. IV
we present various monogamy inequalities for bipartite quantum entanglement
and partial disorder in multipartite states. In Sec. V are the conclusions.

\section{Information-theoretic entanglement measure and multipartite
concurrence}

Consider a pure state $|\psi _{N}\rangle $ of $N$ qubits, labeled as $%
1,2,\cdots ,N$; generally we can write $|\psi _{N}\rangle =|\psi \rangle
_{1}\otimes \cdots \otimes |\psi \rangle _{M}$, where $|\psi \rangle _{m}$
are non-product pure states, $m=1,2,\cdots ,M$, and the qubits of different $%
|\psi \rangle _{m}$ have no intersection. Here, we consider the nontrivial
case $M=1$, i.e. $|\psi _{N}\rangle $ itself is not a product pure state.
Any bipartite partition $\mathcal{P}=\mathcal{A}|\mathcal{B}$ will divide
all the qubits into two subsets $\mathcal{A}$ and $\mathcal{B}$. We denote
the number of qubits contained in $\mathcal{A}$, $\mathcal{B}$ as $|\mathcal{%
A}|$ and $|\mathcal{B}|$ respectively. For $N\in even$, there are two
different kinds of bipartite partitions $\mathcal{P}_{\mathcal{I}}$ and $%
\mathcal{P}_{\mathcal{II}}$. If $|\mathcal{A}|,|\mathcal{B}|\in odd$, $%
\mathcal{P}\in \mathcal{P}_{\mathcal{I}}$, otherwise if $|\mathcal{A}|,|%
\mathcal{B}|\in even$, $\mathcal{P}\in \mathcal{P}_{\mathcal{II}}$. Using
the linear entropy \cite{linear entropy}, the mutual information between $%
\mathcal{A}$ and $\mathcal{B}$ is $S_{\mathcal{A}|\mathcal{B}}=S_{\mathcal{A}%
}+S_{\mathcal{B}}-S_{\mathcal{AB}}$, where $S_{Y}=1-Tr\rho _{Y}^{2}$, $Y=%
\mathcal{A},\mathcal{B},\mathcal{AB}$. Since $|\psi _{N}\rangle $ is a pure
state, we can write $S_{\mathcal{A}|\mathcal{B}}=2(1-Tr\rho _{\mathcal{A}%
}^{2})=2(1-Tr\rho _{\mathcal{B}}^{2})$, where $\rho _{\mathcal{A}}$ and $%
\rho _{\mathcal{B}}$ are the reduced density matrices. The
information-theoretic measure of multi-qubit entanglement is defined as \cite%
{ITMGE}
\begin{equation}
\mathcal{E}_{12\cdots N}=\sum\limits_{\mathcal{P}\in \mathcal{P_{I}}}S_{%
\mathcal{P}}-\sum\limits_{\mathcal{P}\in \mathcal{P_{II}}}S_{\mathcal{P}}
\end{equation}%
where $S_{\mathcal{P}}$ is the mutual information for bipartite partition $%
\mathcal{P}$. It has shown been that $\mathcal{E}_{12\cdots N}$ is a
polynomial SLOCC invariant, and is unchanged under permutations of qubits,
i.e., it represents a collective property of all the $N$ qubits \cite{ITMGE}%
. In the following, we will connect $\mathcal{E}_{12\cdots N}$ with a single
factorizable observable, and as a product, we prove that $\mathcal{E}%
_{12\cdots N}\geq 0$ is satisfied for pure states of an arbitrary even
number qubits.

We can see that the bipartite mutual information $S_{\mathcal{P}}$ is a
quadratic polynomial function of the elements of the reduced density
matrices.\ For any $n$th-degree polynomial function $f$ of the elements of a
density matrix $\rho $, one could always find an observable $A$ on $n$
copies of $\rho ,$ without quantum state tomography \cite{MPFS} $f(\rho
)=Tr(A\rho ^{\otimes n})$. Several important experimental schemes for
measuring nonlinear properties of quantum states through multiple copies of
quantum states have been also proposed \cite{Horo1,Horo2}. In particular,
Mintert and co-workers show that it is possible to measure the multipartite
concurrence of pure states by detecting a single factorizable observable on
two copies of the composite states \cite{Mintert1,Mintert0,Mintert3}.

Given an N-partite pure state
\begin{equation*}
|\psi _{N}\rangle \in \mathcal{H=}\bigotimes\limits_{i=1}^{N}\mathcal{H}_{i}
\end{equation*}
where $\mathcal{H}_{i}$ is the Hilbert space associated to the $i$th
subsystem and $N\in even$. We can define two operators $P_{+}^{i}$ and $%
P_{-}^{i}$ as the projectors onto the symmetric and antisymmetric subspace
of the Hilbert space $\mathcal{H}_{i}\otimes \mathcal{H}_{i}$, which
describes the two copies of the $i$th subsystem. Using these two kinds of
projectors, a set of factorizable observables are introduced \cite%
{Mintert1,Mintert0}
\begin{equation}
A_{s_{1}s_{2}\cdots s_{N}}=P_{s_{1}}^{1}\otimes P_{s_{2}}^{2}\otimes \cdots
\otimes P_{s_{N}}^{N}
\end{equation}%
where $s_{1,}s_{2},\cdots ,s_{N}=+,-$. The expectation value of such an
observable $A_{s_{1}s_{2}\cdots s_{N}}$ with respect to $|\psi _{N}\rangle
\otimes $ $|\psi _{N}\rangle $ is represented as
\begin{equation}
\left\langle A_{s_{1}s_{2}\cdots s_{N}}\right\rangle =\left\langle \psi
_{N}\right\vert \left\langle \psi _{N}\right\vert A_{s_{1}s_{2}\cdots
s_{N}}|\psi _{N}\rangle |\psi _{N}\rangle
\end{equation}%
For simplicity, we denote the number of $-$ in $\left\{ s_{1,}s_{2},\cdots
,s_{N}\right\} $ as $\mathcal{N}_{a}(\left\{ s_{1,}s_{2},\cdots
,s_{N}\right\} )$. If $\mathcal{N}_{a}(\left\{ s_{1,}s_{2},\cdots
,s_{N}\right\} )$ $\in odd$, then $\left\langle A_{s_{1}s_{2}\cdots
s_{N}}\right\rangle =0$. For general $N$-party mixed states described by the
density matrix $\rho $, the purity of $\rho $ can be expressed through the
expectation values of different observables \cite{Mintert0,Mintert1}
\begin{equation}
Tr\rho ^{2}=1-2\sum_{\mathcal{N}_{a}(\left\{ s_{1,}s_{2},\cdots
,s_{N}\right\} )\in odd}Tr(A_{s_{1}s_{2}\cdots s_{N}}\rho ^{\otimes 2})
\end{equation}%
With this knowledge in hand, we present the following lemma about the
information-theoretic multi-qubit entanglement measure $\mathcal{E}%
_{12\cdots N}$ for general even number $N$.

\textbf{\emph{Lemma 1}} \emph{For pure states of even number }$N$\emph{\
qubits, the information-theoretic measure of multi-qubit entanglement }$%
\mathcal{E}_{12\cdots N}$\emph{\ defined in Eq.(1) satisfies}
\begin{equation}
\mathcal{E}_{12\cdots N}=2^{N}\left\langle P_{-}^{1}\otimes P_{-}^{2}\otimes
\cdots \otimes P_{-}^{N}\right\rangle \geq 0
\end{equation}%
\emph{Proof. }The mutual information of a bipartite partition $\mathcal{P}=%
\mathcal{A}|\mathcal{B}$ is\ $S_{\mathcal{A}|\mathcal{B}}=2(1-Tr\rho _{%
\mathcal{A}}^{2})$. We denote the index set $\mathcal{A}$ and $\mathcal{B}$
as $\left\{ a_{1},a_{2},\cdots ,a_{_{|\mathcal{A}|}}\right\} $, $\left\{
b_{1},b_{2},\cdots ,b_{_{|\mathcal{B}|}}\right\} $ respectively. According
to Eq.(4), $S_{\mathcal{A}|\mathcal{B}}$ can be expressed as follows
\begin{equation*}
S_{\mathcal{A}|\mathcal{B}}=4\sum_{\mathcal{N}_{a}(\left\{ s_{i}|i\in
\mathcal{A}\right\} )\in odd,\mathcal{N}_{a}(\left\{ s_{i}|i\in \mathcal{B}%
\right\} )\in odd}\left\langle A_{s_{1}s_{2}\cdots s_{N}}\right\rangle
\end{equation*}%
where $A_{s_{1}s_{2}\cdots s_{N}}$ are the factorizable observables defined
in Eq.(2) on $\mathcal{H}\otimes \mathcal{H}$. In the above derivation, we
have used two important properties that $P_{+}^{i}+P_{-}^{i}=I_{i}$ and $%
\left\langle A_{s_{1}s_{2}\cdots s_{N}}\right\rangle =0$ when $\mathcal{N}%
_{a}(\left\{ s_{1,}s_{2},\cdots ,s_{N}\right\} )$ $\in $ $odd$ for pure
states. If the expectation value of some observable $\left\langle
A_{s_{1}s_{2}\cdots s_{N}}\right\rangle $ contributes to the mutual
information for some bipartite partition $\mathcal{P}=\mathcal{A}|\mathcal{%
B\in P_{II}}$, there must exist one minimum index, denoted as $\mathcal{X}$
that $s_{_{\mathcal{X}}}=+$. If $\mathcal{X\in A}$, $\left\langle
A_{s_{1}s_{2}\cdots s_{N}}\right\rangle $ will also contribute to the mutual
information for some bipartite partition $\mathcal{P}^{\prime }=\mathcal{A-}%
\left\{ \mathcal{X}\right\} |\mathcal{B+}\left\{ \mathcal{X}\right\}
\mathcal{\in P_{I}}$. Otherwise if $\mathcal{X\in B}$, $\left\langle
A_{s_{1}s_{2}\cdots s_{N}}\right\rangle $ will contribute to the mutual
information for some bipartite partition $\mathcal{P}^{\prime }=\mathcal{A+}%
\left\{ \mathcal{X}\right\} |\mathcal{B-}\left\{ \mathcal{X}\right\}
\mathcal{\in P_{I}}$. Conversely, if the expectation value of some
observable $\left\langle A_{s_{1}s_{2}\cdots s_{N}}\right\rangle $
contributes to the mutual information for some bipartite partition $\mathcal{%
P\in P_{I}}$, as long as $A_{s_{1}s_{2}\cdots s_{N}}\neq P_{-}^{1}\otimes
P_{-}^{2}\otimes \cdots \otimes P_{-}^{N}$, it will also contribute to the
mutual information for some bipartite partition $\mathcal{P}^{\prime }%
\mathcal{\in P_{II}}$. According to this corresponding relation, and noting
that the number of bipartite partitions in $\mathcal{P_{I}}$, $\mathcal{%
P_{II}}$ is $2^{N-2}$ and $2^{N-2}-1$\ respectively, we can easily obtain
that $\mathcal{E}_{12\cdots N}=\sum\limits_{\mathcal{P}\in \mathcal{P_{I}}%
}S_{\mathcal{P}}-\sum\limits_{\mathcal{P}\in \mathcal{P_{II}}}S_{\mathcal{P}%
}=2^{N}\left\langle P_{-}^{1}\otimes P_{-}^{2}\otimes \cdots \otimes
P_{-}^{N}\right\rangle $. The expectation value $\left\langle
P_{-}^{1}\otimes P_{-}^{2}\otimes \cdots \otimes P_{-}^{N}\right\rangle $ is
the probability of observing the two copies of each individual subsystem in
an antisymmetric state, which is always a non-negative value. Thus we can
obtain that the entanglement measure $\mathcal{E}_{12\cdots N}\geq 0$.
\textit{It should also be emphasized that the above proof is independent on
the dimensions of individual subsystems }$H_{i}$\textit{, thus the result
that }$\sum_{\mathcal{P}\in \mathcal{P_{I}}}S_{\mathcal{P}}-\sum_{\mathcal{P}%
\in \mathcal{P_{II}}}S_{\mathcal{P}}=2^{N}\left\langle P_{-}^{1}\otimes
P_{-}^{2}\otimes \cdots \otimes P_{-}^{N}\right\rangle \geq 0$\textit{\ is
applicable for arbitrary }$d_{1}\otimes d_{2}\otimes \cdots \otimes d_{N}$%
\textit{\ composite systems, with }$N\in even$\textit{.} $\square $

Together with the other properties of $\mathcal{E}_{12\cdots N}$ presented
in Refs. \cite{ITMGE}, including invariant under local unitary operations
and SLOCC operations, we can easily verify that $\mathcal{E}_{12\cdots N}$
does not increase under local quantum operations assisted with classical
communication (LOCC), thus satisfies all the necessary conditions for a
natural entanglement measure of pure states for general even number $N$. The
result in lemma 1 establishes a direct connection between our
information-theoretic multi-qubit entanglement measure and the class of
concurrence proposed by Mintert and co-workers \cite{Mintert1,Mintert0}. As
pointed out in \cite{Mintert0}, the special concurrence defined through $%
16\left\langle P_{-}^{1}\otimes P_{-}^{2}\otimes P_{-}^{3}\otimes
P_{-}^{4}\right\rangle $ for pure states of four qubits can effectively
characterize separability properties independent of any pairing of
subsystems, i.e. vanishes for any state where at least one subsystem is
uncorrelated with all other system components. This also supports our
proposed information-theoretic measure for multi-qubit entanglement.
Moreover, our measure comes from the information-theoretic viewpoint, thus
the relation in Eq. (5) shows that Mintert concurrence also reflects the
information nature of multi-qubit entanglement, which may help to understand
the nature and structure of entanglement in multipartite entangled states.

We should note that lemma 1 is only valid for an even number of subsystems.
Nevertheless, from the physical point of view, there is nothing
fundamentally different, as for what concerns entanglement, between systems
with an even number of subsystems and those with an odd number. This kind of
limitation stems from the mathematics foundation of the
information-theoretic measure for multi-qubit entanglement. If $N$ is an odd
number, two different kinds of bipartite partitions $\mathcal{P}_{\mathcal{I}%
}$ and $\mathcal{P}_{\mathcal{II}}$ do not exist anymore, while $%
\left\langle P_{-}^{1}\otimes P_{-}^{2}\otimes \cdots \otimes
P_{-}^{N}\right\rangle $ will always vanish in coincidence.

\section{Compatibility conditions for multipartite states}

The compatibility problem can be formulated as in \cite%
{Butterley,Hall0610031}: Given states of all proper subsystems of a
multipartite quantum system, what are the necessary and sufficient
conditions for these subsystem states to be compatible with a single entire
system? A number of partial results have been obtained through different
ideas and approaches. Higuchi et al. find the necessary and sufficient
conditions for the possible one-qubit reduced states of a pure multi-qubit
state\cite{Higuchi0}. This result is then generalized for $3\otimes 3\otimes
3$ \cite{Higuchi0309186} and for $2\otimes 2\otimes 4$ \cite{Bravyi}
systems. An interesting necessary condition for an odd n-party state has
also been proposed in \cite{Butterley,Hall2005}. In \cite%
{Christandl0409016,Christandl0511029}, Christandl et al. establish an
connection between the compatibility problem and the representation theory
of the symmetric group. Most recently, Hall has attack the compatibility
problem from a novel angel, i.e. utilizing the ideas of convexity \cite%
{Hall0610031}. However, there are very few necessary criteria for the
general form of the compatibility problem. In this section, we will derive a
set of compatibility conditions for general $d_{1}\otimes d_{2}\otimes
\cdots \otimes d_{N}$ multipartite states from the relationship between the
information-theoretic entanglement measure and multipartite concurrence.

In Eq. (5), the invariant $\mathcal{E}_{12\cdots N}$ is directly determined
by the properties of reduced density matrices, therefore based on the above
lemma, it is easy for us to get a simple necessary condition for the
compatibility problem of arbitrary $d_{1}\otimes d_{2}\otimes \cdots \otimes
d_{N}$ composite systems from multi-qubit entanglement measures.

\textbf{Theorem 1} \emph{Given a set of density matrices $\left\{ \rho
_{1},\cdots \rho _{N},\rho _{12,}\cdots \rho _{N-1N},\cdots ,\rho _{12\cdots
N-1},\cdots,\rho _{2\cdots N-1N}\right\} ,$ if they come from one common
N-party pure states $|\psi _{N}\rangle$, the following inequality should be
satisfied}
\begin{equation}
\sum_{|\mathcal{A}|\in odd}Tr\rho _{\mathcal{A}}^{2}-\sum_{|\mathcal{A}|\in
even}Tr\rho _{\mathcal{A}}^{2}\leq 1
\end{equation}%
where
\begin{equation*}
\mathcal{A}\subseteq \mathcal{N}=\left\{ 1,2,\cdots ,N\right\} \text{ and }%
\mathcal{A}\neq \varnothing
\end{equation*}

\emph{Proof. }First, we assume that $N\in even$, using the analysis in lemma
1, we can write the mutual information for a bipartite partition $\mathcal{P}%
=\mathcal{A}|\mathcal{B}$, with $\mathcal{A},\mathcal{B}$ $\neq \varnothing $%
, as $S_{\mathcal{A}|\mathcal{B}}=2-Tr\rho _{\mathcal{A}}^{2}-Tr\rho _{%
\mathcal{B}}^{2}$. Therefore, $\sum_{\mathcal{P}\in \mathcal{P_{I}}}S_{%
\mathcal{P}}=2\cdot 2^{N-2}-\sum_{|\mathcal{A}|\in odd}Tr\rho _{\mathcal{A}%
}^{2}$ and $\sum_{\mathcal{P}\in \mathcal{P_{II}}}S_{\mathcal{P}}=2\cdot
(2^{N-2}-1)-\sum_{|\mathcal{A}|\in even,\text{ }\mathcal{A}\neq \mathcal{N}%
}Tr\rho _{\mathcal{A}}^{2}$, which results in that $\mathcal{E}_{12\cdots
N}=2-\sum_{|\mathcal{A}|\in odd}Tr\rho _{\mathcal{A}}^{2}+\sum_{|\mathcal{A}%
|\in even,\text{ }\mathcal{A}\neq \mathcal{N}}Tr\rho _{\mathcal{A}}^{2}$.
According to the result in Eq.(5) that $\mathcal{E}_{12\cdots N}\geq 0$, and
note that if $\mathcal{A}=\mathcal{N}$, $Tr\rho _{\mathcal{A}}^{2}=1,$ we
prove that the above necessary condition is satisfied for general even
number $N$. If $N\in odd$, it is obvious that $\sum_{|\mathcal{A}|\in
odd}Tr\rho _{\mathcal{A}}^{2}-\sum_{|\mathcal{A}|\in even}Tr\rho _{\mathcal{A%
}}^{2}=1$, thus we finish the proof of theorem 1. $\square $

The equality in theorem 1 will be satisfied when $\mathcal{E}_{12\cdots N}=0$%
, e.g. for general $N-$qubit $W$ states $\left\vert W_{N}\right\rangle =%
\frac{1}{\sqrt{N}}(\left\vert 0\cdots 01\right\rangle +\left\vert 0\cdots
010\right\rangle +\cdots +\left\vert 10\cdots 0\right\rangle )$. In the
above derivation of theorem 1, the necessary condition for the compatibility
problem follows from the fact that $\left\langle P_{-}^{1}\otimes
P_{-}^{2}\otimes \cdots \otimes P_{-}^{N}\right\rangle \geq 0$. However,
there are other similar factorizable observables such as $\left\langle
P_{s_{1}}^{1}\otimes P_{s_{2}}^{2}\otimes \cdots \otimes
P_{s_{N}}^{N}\right\rangle \geq 0$, with $\mathcal{N}_{a}(\left\{
s_{1,}s_{2},\cdots ,s_{N}\right\} )$ $\in even$. Therefore, following the
same idea, we can generalize the above compatibility condition to general
multipartite mixed states.

\textbf{\emph{Theorem 2}}\emph{\ Given an $N$-party state $\rho$, if $N=2k$
is even, the reduced density matrices should satisfy the following
inequalities}
\begin{equation}
\sum_{|\mathcal{A}|\in odd}Tr\rho _{\mathcal{A}}^{2}-\sum_{|\mathcal{A}|\in
even}Tr\rho _{\mathcal{A}}^{2}\leq 1
\end{equation}%
where
\begin{equation*}
\mathcal{A}\subseteq \mathcal{N}=\left\{ 1,2,\cdots ,2k\right\} \text{ and }%
\mathcal{A}\neq \varnothing
\end{equation*}%
\emph{Proof.} We could always find an ancillary subsystem $\mathcal{R}$ and
a pure state $|\psi\rangle_{\mathcal{NR}}$, such that $Tr_\mathcal{R}%
(|\psi\rangle_{\mathcal{NR}}\langle\psi|)=\rho$. In the similar way as the
proof of theorem 1, using the expression of purity for reduced density
matrices based on the expectation values in Eq.(4), and after some
straightforward calculations, we can get $1-\sum_{|\mathcal{A}|\in
odd}Tr\rho _{\mathcal{A}}^{2}+\sum_{|\mathcal{A}|\in even}Tr\rho _{\mathcal{A%
}}^{2}=2^{\left\vert \mathcal{N}\right\vert }\left\langle P_{-}^{1}\otimes
P_{-}^{2}\otimes \cdots \otimes P_{-}^{2k}\otimes \mathbf{P}_{+}^{\mathcal{R}%
}\right\rangle $, where $\mathbf{P}_{+}^{\mathcal{R}}$ is the projector onto
the globally symmetric subspace of the Hilbert space $\mathcal{H}_{\mathcal{R%
}}\otimes \mathcal{H}_{\mathcal{R}}$. Since the expectation value $%
\left\langle P_{-}^{1}\otimes P_{-}^{2}\otimes \cdots \otimes
P_{-}^{2k}\otimes \mathbf{P}_{+}^{\mathcal{R}}\right\rangle $ is always a
non-negative value, we thus prove the necessary conditions in Eq.(7). $%
\square $


The compatibility conditions in theorem 1 and 2 directly come from the
properties of multipartite entanglement measures, this basic idea is much
different from previous works. In addition, as discussed in the beginning of
this section, most known results on the compatibility problem are about
one-party or two-party reduced states. However, our compatibility conditions
are about general density matrices of all proper subsystems of an $N-$party
quantum system, including one-party, two-partite, $\cdots $, and $(N-1)$%
-party states. Therefore, these compatibility conditions will be more
powerful in the situation of general $k-$party reduced states. We could
construct a simple example similar to the one in \cite{Hall0610031} to
explicitly demonstrate the strength of our criteria. Consider the following
states of all proper subsystems of a four-qubit system:%
\begin{eqnarray}
\rho _{1} &=&\rho _{2}=\rho _{3}=\rho _{4}=\left(
\begin{array}{cc}
\frac{2}{3} & 0 \\
0 & \frac{1}{3}%
\end{array}%
\right)   \notag \\
\rho _{12} &=&\rho _{13}=\rho _{14}=\rho _{23}=\rho _{24}=\rho _{34}=\left(
\begin{array}{cccc}
\frac{1}{3} & 0 & 0 & 0 \\
0 & \frac{1}{3} & \frac{1}{3} & 0 \\
0 & \frac{1}{3} & \frac{1}{3} & 0 \\
0 & 0 & 0 & 0%
\end{array}%
\right)   \notag \\
\rho _{123} &=&\rho _{124}=\rho _{134}=\rho _{234}=|W\rangle \langle W|
\notag \\
|W\rangle  &=&\frac{1}{\sqrt{3}}(|001\rangle +|010\rangle +|100\rangle )
\end{eqnarray}%
which are not compatible with an overall four-qubit state $\rho _{1234}$
(not necessary a pure state). In this example, the previous results can only
determine whether one-party and two-party states come from a single
four-qubit state. However, there are no known conditions for three-party
reduced states. If we check the necessary condition in Eq. (7), it can be
seen that $(1\times 4+\frac{5}{9}\times 4)-(\frac{5}{9}\times 6+Tr\rho
_{1234}^{2})\leq 1$ will always not be satisfied. Therefore, theorem 2
confirms the incompatibility of the density matrices in Eq. (8).

\section{Applications of compatibility conditions}

Based on the above compatibility conditions, we could obtain a new kind of
monogamy inequalities for bipartite quantum entanglement \cite{Osborne06}
and partial disorder in multipartite states. It is known that the
entanglement measure for arbitrary dimension bipartite pure states, i.e. the
square of \textit{I-concurrence }\cite{Rungta}, is also relevant to the
purity of marginal density matrices\textit{. }For a bipartite pure state $%
\left\vert \Psi _{AB}\right\rangle $, the square of I-concurrence is defined
as
\begin{equation}
C_{A|B}^{2}=2(1-Tr\rho _{A}^{2})=2(1-Tr\rho _{B}^{2})
\end{equation}%
where $\rho _{A}$ and $\rho _{B}$ are reduced density matrices. Therefore,
according to theorem 1 and 2, we can derive a class of monogamy inequalities
\cite{Osborne06} for bipartite entanglement in general multipartite pure
states as follows.

\textbf{\emph{\ Corollary 1 }}\emph{Given an N-party pure states }$|\psi
_{N}\rangle $\emph{, there exist \ }%
\begin{equation}
\sum\limits_{|\mathcal{A}|\in odd}C_{\mathcal{A}|\mathcal{N-A}}^{2}\geq
\sum\limits_{|\mathcal{A}|\in even}C_{\mathcal{A}|\mathcal{N-A}}^{2}
\end{equation}%
\emph{where }$I$\emph{\ }$=\left\{ i_{1},i_{2},\cdots ,i_{2k}\right\}
\subseteq \mathcal{N}$\emph{\ and }$A\subseteq I$\textit{. Here, we assume
that }$C_{\mathcal{A}|\mathcal{N-A}}^{2}=0$\textit{\ when }$A=\varnothing $%
\emph{\ or }$\mathcal{N} $\emph{.}

The above monogamy inequalities put new constrains on the distributed
entanglement. It is valid not only for multipartite states of qubits, but
also for arbitrary dimensions. The monogamous nature of entanglement is much
relevant to quantum cryptography \cite{Osborne06}. In the context of
condensed matter physics, the monogamy property gives rise to some
interesting effects, e.g. frustration in quantum spin systems. Therefore,
these monogamy inequalities for bipartite entanglement may be valuable in
many-body physics.

As another explicit application of our results, we consider a general $%
d_{1}\otimes d_{2}$ bipartite mixed state $\rho _{12}$. The disorder of a
quantum state described by the density matrix $\rho $ can be characterized
by the mixedness $D(\rho )=1-Tr\rho ^{2}$ \cite{mixedness}, in which we
neglect the normalization factor for simplicity. When $\rho $ is a maximally
mixed state, $D(\rho )$ takes the maximum value. According to theorem 2, it
is obvious that $Tr\rho _{1}^{2}+Tr\rho _{2}^{2}-Tr\rho _{12}^{2}\leq 1$.
Therefore, we can obtain a relation between the global disorder and local
disorder
\begin{equation}
D(\rho _{12})\leq D(\rho _{1})+D(\rho _{2})
\end{equation}%
This inequality demonstrates that the global disorder is always no larger
than the sum of local disorder, which is an analog to the subadditivity of
von Neumann entropy $S(A,B)\leq S(A)+S(B)$\cite{Christandl0409016}. However,
it is difficult to generalize this subadditivity based on von Neumann
entropy to arbitrary multipartite mixed states. If we adopt $D(\rho )$ as
the measure of disorder and following the result in theorem 2, it is
possible to achieve more general relations between global and local
disorder. For example, for a general $d_{1}\otimes d_{2}$ $\otimes
d_{3}\otimes d_{4}$ mixed state $\rho _{1234}$, there exists the following
relation between globally and locally disorder
\begin{equation}
D(\rho _{1234})+\sum_{i,j=1}^{4}D(\rho _{ij})\leq \sum_{i=1}^{4}D(\rho
_{i})+\sum_{i,j,k=1}^{4}D(\rho _{ijk})
\end{equation}

It is well known that many multipartite entanglement measures are polynomial
invariants. In this paper, we link two such quantities defined from
completely different viewpoints, the extension of which will help to clarify
and unify the understanding of multipartite entanglement. Similarly, the
connections of other multipartite entanglement measures will also lead to
interesting results about general multipartite states.

\section{Conclusion}

In summary, we present necessary conditions for general $m$-party subsystem
states to be the reduced states of a single multipartite state of arbitrary $%
d_{1}\otimes d_{2}\otimes \cdots \otimes d_{N}$ composite systems. Our
method is based on the properties of multipartite entanglement measures,
rather than directly investigating the reduced density matrices as in
previous work. These results clearly demonstrate the close connection
between multipartite entanglement and the general compatibility problem. As
a consequence, we get some interesting monogamy inequalities for bipartite
entanglement and partial disorder in general multipartite states.

\emph{Acknowledgments} This work was funded by the National Fundamental
Research Program 2006CB921900, Grant No. NCET-04-0587, the Innovation funds
from Chinese Academy of Sciences, and National Natural Science Foundation of
China (Grants Nos. 60621064, 10574126).

\end{document}